\documentclass[aps, prl, twocolumn,tightenlines]{revtex4}
\usepackage{epsfig}
\begin{document}

\title{Coherent Anti-Stokes Raman Scattering Heterodyne Interferometry}
\author{J.S. Bredfeldt}
\author{D.L. Marks}
\author{C. Vinegoni}
\author{S. Hambir}
\affiliation{Beckman Institute for Advanced Science and
Technology, University of Illinois at Urbana-Champaign}
\author{S.A. Boppart}\email{boppart@uiuc.edu}
\affiliation{Department of Electrical and Computer Engineering,
Bioengineering Program, College of Medicine, Beckman Institute for
Advanced Science and Technology, University of Illinois at
Urbana-Champaign, 405 North Mathews Avenue, Urbana, IL 61801}
\date{\today}

%%%%%%%%%%%%%%%%%%%%%%%%%%%%
% ABSTRACT
%%%%%%%%%%%%%%%%%%%%%%%%%%%%%

\begin{abstract}
A new interferometric technique is demonstrated for measuring
Coherent Anti-Stokes Raman Scattering (CARS) signals. Two
forward-directed CARS signals are generated, one in each arm of an
interferometer. The deterministic nature of the CARS process
allows for these two signals, when spatially and temporally
overlapped, to interfere with one another. Heterodyne detection
can therefore be employed to increase the sensitivity in CARS
signal detection. In addition, nonlinear CARS interferometry will
facilitate the use of this spectroscopic technique for molecular
contrast in Optical Coherence Tomography (OCT).
\\

\end{abstract}

 \maketitle

%%%%%%%%%%%%%%%%%%%%%%%%%%%%%%%%%
%  PAPER
%%%%%%%%%%%%%%%%%%%%%%%%%%%%%%%%
Optical Coherence Tomography (OCT) is an interferometric optical
imaging technique capable of imaging tissue microstructure at near
histological resolutions \cite{boppart1998}. Unfortunately, the
linear scattering properties of pathological tissue probed by OCT
are often morphologically and/or optically similar to normal
tissue. To address this problem novel contrast methods for OCT
have been recently developed, such as spectroscopic OCT
\cite{morgner2000}, a pump and probe technique \cite{rao2003}, and
the use of engineered microspheres \cite{lee2003} or microbubbles
\cite{barton2002}.

Spectroscopic OCT (SOCT) measures the spectral absorption from
tissues by measuring the spectral differences between the source
and the backscattered interference signal to provide information
about the properties of the scatterers in the sample. However,
this technique is limited to the identification of scatterers that
have absorption within the bandwidth of the optical source. A
different method to obtain contrast enhanced OCT employs
engineered microsphere contrast agents, which can be targeted to
cell receptors thereby providing molecular specific contrast
\cite{lee2003}. A drawback to this technique, however, is that the
contrast agents may negatively impact the biology under
investigation.   We present a new method for achieving enhanced
OCT contrast, exploiting the inherent vibrational frequency
differences between molecular bonds within the tissues. The
spectroscopic technique that is employed to detect these
vibrational differences is Coherent Anti-Stokes Raman Scattering
(CARS).

CARS is a well-known spectroscopic technique that has recently
received significant attention for its applications to scanning
microscopy. In CARS spectroscopy, the frequencies of two incident
lasers,  $\omega_p$ (pump) and  $\omega_s$ (Stokes), are chosen
such that the difference  $\omega_p - \omega_s = \omega_v$ is
equal to a Raman-active vibrational mode of the molecule under
study \cite{dermtroeder1998}. CARS is a non-linear, four-wave
mixing process. Furthermore, the CARS field is a result of the
interaction between four photons and is generated in the
phase-matching direction at the anti-Stokes frequency $\omega_{AS}
= 2 \omega_p - \omega_s$, implying that the CARS signal intensity
is linearly dependent on the Stokes field intensity and
quadratically dependent on the pump field intensity. Note that, in
addition to the CARS signal, a broadband non-resonant background
is always present, limiting the vibrational contrast achieved in
CARS microscopy. However, CARS is a coherent process, with the
phase of the anti-Stokes field deterministically related to the
phase of the excitation field. Therefore, constructive
interference of the anti-Stokes field causes the CARS signal to be
significantly larger than the spontaneous Raman signal, given the
same average excitation power \cite{cheng2002bis}. All these
characteristics have allowed CARS to be successfully employed to
provide vibrational contrast in scanning microscopy
\cite{cheng2002bis,duncan1982,zumbusch1999,wurpel2002,hashimoto2000}.

CARS scanning microscopy generally involves scanning overlapped
and tightly focused pump and Stokes lasers through a sample while
measuring the anti-Stokes signal amplitude point by point. The
first CARS microscope \cite{duncan1982} utilized non-collinear
pump and Stokes visible lasers to demonstrate microscopic imaging
of the spatial distribution of deuterated molecular bonds in a
sample of onion skin. Tightly focused, collinear near-infrared
pump and Stokes pulses were used \cite{zumbusch1999} to achieve
improved background signal suppression, and three-dimensional
sectioning in living cells. In each of these CARS microscopy
techniques, the anti-Stokes photons are counted in order to
estimate the density of the Raman scatterers and/or Raman
susceptibility magnitude in the focal volume of the microscope.
However, the spectral phase information is lost in this process
and can only be inferred. In this Letter we propose and
demonstrate a new CARS interferometric technique called Nonlinear
Interferometric Vibrational Imaging (NIVI) with the capability for
heterodyne detection and the possibility to obtain a full
reconstruction of the magnitude and phase of the sample Raman
susceptibility.

\begin{figure}
\epsfig{figure=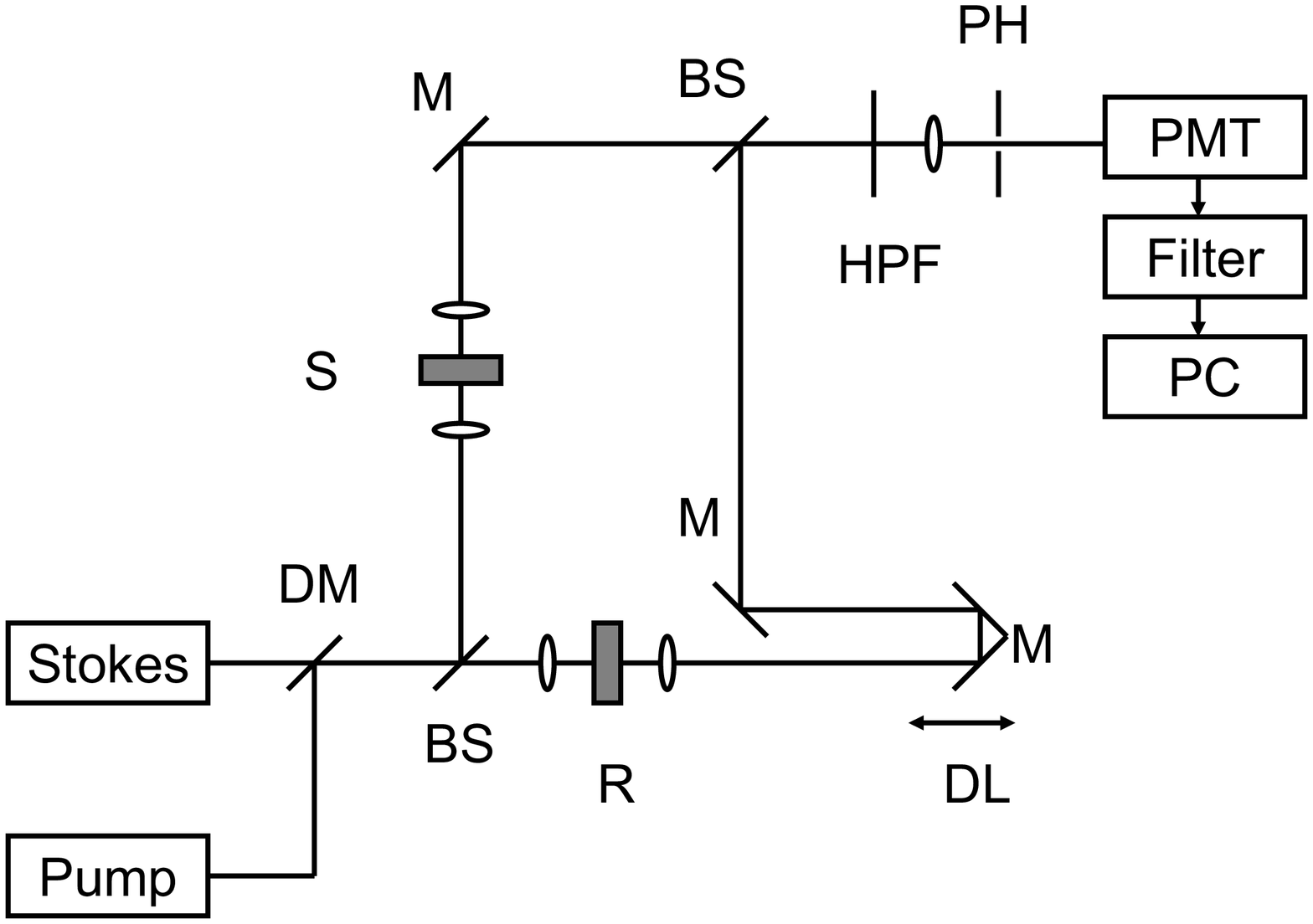,width=90mm,height=69mm}
\caption{Schematic of the interferometric CARS measurement system.
Benzene was used for both the reference and sample materials in
this study. Abbreviations: S, sample; R, reference; DM, dichroic
mirror; BS, beamsplitter; M, mirror; HPF, high-pass-filter; PH,
pin-hole; DL, delay line. \label{setup}}
\end{figure}

The CARS interferometer described in this paper is presented in
Fig.\ref{setup}.
An excitation field consisting of two overlapped pulses centered
at the pump and Stokes wavelengths is divided by a beamsplitter
into two separate interferometer paths, which are referred to in
Fig.\ref{setup} as arm "S" (or sample arm) and arm "R" (or
reference arm). A sample of a molecule is placed into each arm
into which the split excitation fields are focused.  If the
frequency difference between the pump and Stokes pulses is tuned
to a Raman active vibrational mode present in both sample S and
sample R, an anti-Stokes signal is generated in each arm of the
interferometer. Because the anti-Stokes pulse phase is
deterministically related to the phase of the pump and the Stokes
pulses, the anti-Stokes fields are coherent with the excitation
fields and also with each other. It follows (Feynman principle)
that when these anti-Stokes fields are temporally and spatially
overlapped, interference can be observed.

In our current setup the CARS is stimulated by a laser system
similar to that employed by \cite{zumbusch1999}. A diode pumped
frequency doubled Nd:YVO$_4$ laser is used to pump a mode-locked
Ti:sapphire oscillator operating at a center wavelength of 807 nm,
with a bandwidth of 30 nm, a repetition rate of 82 MHz, and an
average power of 300 mW. These pulses seed a regenerative chirped
pulse amplifier (Coherent, RegA 9000) producing approximately 70
fs, 5 µJ pulses at a repetition rate of 250 kHz with an average
power of 1.25 W. Ten percent of this average power is used as the
pump beam while the remaining power is directed to an optical
parametric amplifier (Coherent, OPA 9400) which generates a 4 mW
average power Stokes beam, tunable from 400-1200 nm.

The pump and Stokes pulses, at 807 and 1072 nm respectively, are
used to excite the strong, isolated Raman-active vibrational mode
of benzene at 3063 cm$^{-1}$. As shown in Fig.\ref{setup}, these
pulses are collinearly overlapped using a dichroic mirror and
split with a 50:50 ultrafast beamsplitter into arms S and R. In
each arm, a 30 mm focal length, 12 mm diameter achromatic lens is
used to focus the pump and Stokes beams into a quartz cuvette
filled with benzene. The anti-Stokes signals generated in each arm
are collected in the collinear phase matching direction using 30
mm focal length, 12 mm diameter singlet lenses. The two
anti-Stokes pulses are overlapped in time by adjusting the
relative delay and in space by adjusting the position on a second
beamsplitter.  A high-pass filter at 742 nm eliminates the
remaining excitation light. The filtered anti-Stokes signal is
spatially filtered through a 50 µm diameter pin hole. The relative
delay is scanned by a computer-controlled single axis translation
stage at a constant rate in arm R, and the CARS signal intensity
is measured with a photomultiplier tube PMT (Hamamatsu, HC 123).
Lastly, the signal from the PMT is filtered with a low-pass
anti-aliasing filter and sampled with a PC based data acquisition
system.
\begin{figure}
\epsfig{figure=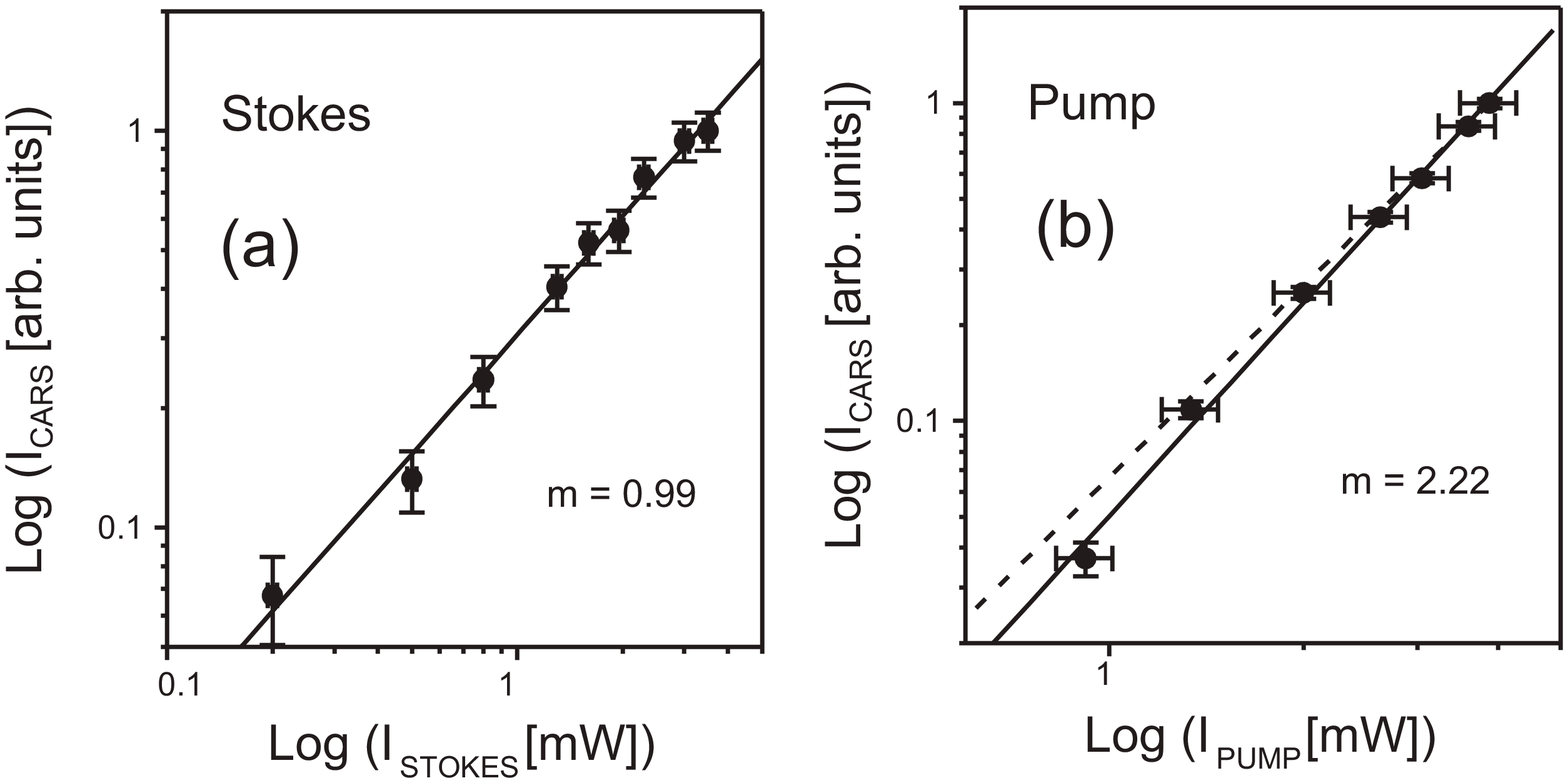,width=90mm,height=69mm}
\caption{Intensity of the CARS signal as a function of (a) the
intensity of the Stokes field and (b) the intensity of the pump
field. Both figures are log-log plots and the solid lines
represent curve fitting. The dotted line of Fig.\ref{fwm}(b) has a
slope of 2. The parameter m is equal to the angular coefficient of
the solid lines. \label{fwm}}
\end{figure}

Figures \ref{fwm}(a) and \ref{fwm}(b) show the observed
relationship between the CARS and the Stokes intensity (pump
intensity fixed) and the CARS and the pump intensity (Stokes
intensity fixed), respectively. The solid lines represent linear
fits of the experimental data. In agreement with theory, the slope
of the fitted lines verifies the linear relationship between the
anti-Stokes and the Stokes intensities and the quadratic
relationship between the anti-Stokes and the pump intensities. Our
signal is therefore a result of a four-wave mixing process.
Moreover, this process is CARS resonance because the anti-Stokes
power is maximized when the Stokes wavelength is tuned to
resonance with the Raman-active benzene vibrational mode.

\begin{figure}
\epsfig{figure=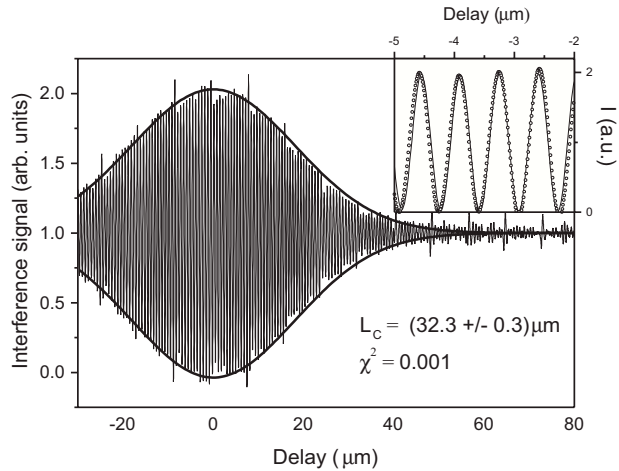,width=90mm,height=69mm} \caption{CARS
interferogram detected at the beamsplitter BS and  produced as the
pathlength of arm R is scanned. The modulus of the degree of the
coherence function is used to fit to the envelope of the
interferogram. The inset shows an enlarged version of the
interference pattern and its fit by the real part of the degree of
coherence function. L$_C$ is the coherence length of the CARS
pulse. \label{interference}}
\end{figure}

Fig.\ref{interference} contains the measured interferogram
resulting from the interference between the two anti-Stokes
signals at the beamsplitter BS. The real part and the modulus of
the coherence function, for the case of a Gaussian spectral
distribution, are used to fit the experimental data (interferogram
and envelope respectively). The resulting coherence length L$_C$,
or the axial resolution of the interferometric CARS measurement
technique, is found to be equal to (32.3 +/- 0.3) $\mu$m  (reduced
$\chi^2$  = 0.001).

The inset of Fig.\ref{interference} shows an enlarged version of
the interferogram fringes. Open circles correspond to the
experimentally measured data and the solid line represents a fit
of the data. This result indicates that two anti-Stokes signals
generated in separate samples can be demodulated
interferometrically, where the amplitude of the fringe envelope
gives information about the concentration of the scatterers in the
focal volume of the sample objective lens. The presence of
interference clearly demonstrates the potential of CARS as a
promising technique for providing molecular contrast for OCT-like
interferometric imaging systems. In fact, the presence of the
interference indicates that similar Raman-active vibrational
frequencies are present in both the reference and the sample arm
at the same path length from the detector. The ``fingerprint''
nature of Raman spectroscopy, combined with the possibility to
switch between different samples in arm R, could permit selective
detection and imaging, within the above mentioned axial
resolution, of different molecular species present in the sample
S.

Note finally that the interferometric detection scheme could
provide numerous advantages over traditional photon counting CARS
microscopy. Interfering a weak CARS signal with another strong
CARS signal, can provide heterodyne sensitivity for improved S/N
ratio. Moreover, with full knowledge of the excitation pulses and
of the CARS interferogram, the spectral amplitude and the phase of
the CARS pulse can be measured. A complete reconstruction of the
Raman susceptibility may then be attained allowing for a more
accurate molecular identification \cite{marks2003}.

In conclusion, we have described a new technique for CARS
measurement that relies on the deterministic nature of the CARS
process. The interference between two CARS signals, generated in
separate samples, was observed allowing for heterodyne detection.
This result is extremely promising for the development of a new
molecular imaging technique (NIVI) based on non-linear,
low-coherence interferometry. While this demonstration used
forward CARS, epi-detected CARS \cite{cheng2002bis} is coherent as
well, and is compatible with OCT coherence-ranging systems. CARS
interferometry provides CARS microscopy the advantages of
interferometric detection and provides OCT with molecular-specific
contrast. These advantages could make CARS interferometry a
powerful tool for biological imaging with OCT and for disease
diagnosis at the molecular level.

This research was supported in part by a research grant entitled
``A Nonlinear OCT System for Biomolecular Detection and
Intervention'' from NASA and the National Cancer Institute
(NAS2-02057, SAB). S.A. Boppart's email address is
boppart@uiuc.edu.

%%%%%%%%%%%%%%%%%%%%%%%%%%%%%%%%%
%  REFERBECES
%%%%%%%%%%%%%%%%%%%%%%%%%%%%%%%%
%Preprint quant-ph/0207179 at khttp://xxx.lanl.govl (2002).
%\bibliographystyle{IEEEtran} % use IEEEtran.bst style
%\bibliography{Bibliography}

\end{document}